\documentclass[aps,prl,twocolumn,footinbib,showpacs,superscriptaddress,secnumarabic,floatfix]{revtex4-1}

\usepackage[english]{babel}

\usepackage{amsfonts,amssymb,amsmath}

\usepackage{graphicx}
\usepackage[caption=false]{subfig}
\usepackage{color}
\usepackage[usenames,dvipsnames]{xcolor}
\usepackage{hyperref}
\hypersetup{colorlinks=true,linkcolor=BrickRed,citecolor=NavyBlue,urlcolor=NavyBlue,linktoc=page}

\usepackage{enumerate}
\usepackage{grffile}
\usepackage{balance}
\usepackage{float}

\newcommand{\Tc}{T_\mathrm{\scriptscriptstyle C}}
\newcommand{\Jc}{J_\mathrm{\scriptscriptstyle C}}
\newcommand{\Hc}{H_\mathrm{{\scriptscriptstyle C}}}
\newcommand{\Hct}{H_\mathrm{{\scriptscriptstyle C}2}}
\newcommand{\Jdp}{J_\mathrm{dp}}
\newcommand{\Np}{N_\mathrm{p}}
\newcommand{\Nv}{N_\mathrm{v}}
\newcommand{\fn}{f_\mathrm{n}}
\newcommand{\ffree}{f_\mathrm{free}}
\newcommand{\ffill}{f_\mathrm{fill}}
\newcommand{\Lt}{L_\mathrm{t}}
\newcommand{\ul}{u_\mathrm{l}}
\newcommand{\ut}{u_\mathrm{t}}
\newcommand{\ff}{\rho_\mathrm{ff}}

\begin{document}

\title{Optimization of vortex pinning by nanoparticles using simulations \\ of time-dependent Ginzburg-Landau model}

\author{A.\,E.\,Koshelev}
\affiliation{Materials Science Division, Argonne National Laboratory, 9700 S. Cass Av., Argonne, Il 60639, USA}

\author{I.\,A.\,Sadovskyy}
\affiliation{Materials Science Division, Argonne National Laboratory, 9700 S. Cass Av., Argonne, Il 60639, USA}

\author{C.\,L.\,Phillips}
\affiliation{Mathematics and Computer Science Division, Argonne National Laboratory, 9700 S. Cass Av., Argonne, Il 60639, USA}

\author{A.\,Glatz}
\affiliation{Materials Science Division, Argonne National Laboratory, 9700 S. Cass Av., Argonne, Il 60639, USA}
\affiliation{Department of Physics, Northern Illinois University, DeKalb, Il 60115, USA}

\date{\today}

\begin{abstract}	
Introducing nanoparticles into superconducting materials has emerged as an efficient route to enhance their current-carrying capability. We address the problem of optimizing vortex pinning landscape for randomly distributed metallic spherical inclusions using large-scale numerical simulations of time-dependent Ginzburg-Landau equations. We found the size and density of particles for which the highest critical current is realized in a fixed magnetic field. For each particle size and magnetic field, the critical current reaches a maximum value at a certain particle density, which typically corresponds to 15--23\% of the total volume being replaced by nonsuperconducting material. For fixed diameter, this optimal particle density increases with the magnetic field. Moreover, we found that the optimal particle diameter slowly decreases with the magnetic field from 4.5 to 2.5 coherence lengths at a given temperature. This result shows that pinning landscapes have to be designed for specific applications taking into account relevant magnetic field scales.
\end{abstract}


\maketitle

High-current applications of superconductivity strongly depend on engineering defect microstructures which efficiently suppress the mobility of vortex lines over a wide range of magnetic fields. Introducing self-assembled inclusions into cuprate superconducting materials has been established as a very efficient route to improve their current-carrying capacity. In particular, such inclusions may be prepared in the form of almost spherical particles~\cite{MacManusAPL04,HauganNat04,GutierrezNatMat07,YamasakiSUST08,PolatPhysRevB11,MiuraPhysRevB11,*MiuraSUST13}, nanorods~\cite{GoyalSUST05,*KangSci06}, or combinations of both~\cite{MaiorovNaMat09}. This technology has been implemented in the second-generation superconducting cables based on YBa$_{2}$Cu$_{3}$O$_{7}$ (YBCO) coated conductors~\cite{MalozemoffAnnRevMatRes12}.

At present, improvement of pinning properties is achieved mostly by trial-and-error approach. Rational design of optimal pinning landscapes may be facilitated by large-scale numerical simulations. Real strong-pinning materials always contain many different defects, such as dislocations, twin boundaries, stacking faults, etc. It is very challenging to understand details of vortex dynamics in such complicated pinning landscape. While the ultimate critical-current optimization can be achieved by the constructive combination of different pinning centers, a natural first step is to determine the best pinning configuration for a relatively simple system with only one type of defects. In this article, we explore the case of monodisperse spherical defects with the size of a few coherence lengths which model nanoparticles in coated YBCO conductors. In addition, similar pinning centers in the form of impurity clusters can be introduced by proton irradiation which provides an opportunity for further enhancement of the critical current~\cite{JiaAPL13,MatsuiJAP15}.

The quantitative description of vortex dynamics is a highly nontrivial problem even for a model system. Pinning is a collective phenomenon controlled by the interaction of vortices with pinning centers as well as the flexibility of vortex lines and intervortex interactions. An analytical treatment of this problem is limited to qualitative estimates of critical currents in simple situations, such as weak pinning by high densities of atomic impurities~\cite{LarkinO:1979} or strong pinning by low densities of strong inclusions interacting with a regular vortex lattice~\cite{OvchinnikovI:1991,BlatterGK:2004}, see also reviews~\cite{BlatterFGLV:1994,Brandt:1995,BlatterG:2003,GurevichAnnRevCMP14}. While providing a useful general guidance, qualitative estimates inevitably rely on simplifying assumptions and miss important details.

To improve a quantitative description of pinning mechanisms, vortex dynamics is extensively explored by numerical simulations using several models with various degrees of complexity and realisticity. The minimal approach is to treat the vortices as elastic strings whose dynamics is described by the overdamped equation of motion taking into account interaction with pinning centers and thermal Langevin forces. Such a Langevin-dynamics approach~\cite{ErtasK:1996,OtterloPRL00,BustingorryCD:2007,LuoHu:2007,Koshelev:2011,DobramyslEPJ13} provides a qualitative description of the vortex state for low fields, when the distance between vortices is much larger than the coherence length, and for a low density of pinning centers. This description, however, has several limitations: vortex-vortex and vortex-defect interactions can be treated only approximately, and vortex cuttings and reconnections are completely neglected. Also, these simplified models cannot be used when pinning centers occupy a noticeable fraction of superconductor's volume corresponding to strongest pinning. Therefore, it is desirable to probe the strong-pinning regime by more sophisticated models.

\begin{figure*}[tb]
	\begin{center}
		\subfloat{\hspace{-0.05in} \includegraphics[width=4.5in]{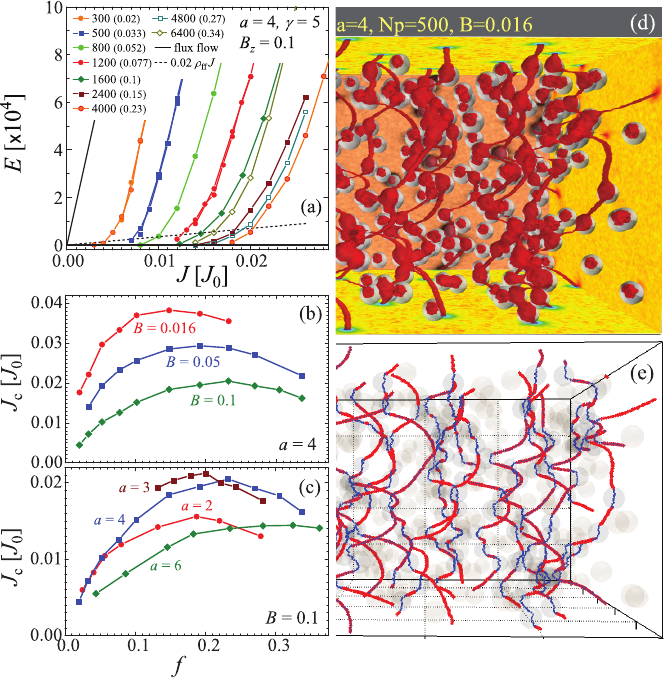} \label{fig:IV}}
		\subfloat{\label{fig:Jc_f_B}}
		\subfloat{\label{fig:Jc_f_a}}
		\subfloat{\label{fig:isosurfaces}}
		\subfloat{\label{fig:vortices}}
	\end{center} \vspace{-0.25in}
	\caption{
		(a)~Current-voltage dependences computed for different numbers of
		particles $\Np$ with diameter $a = 4\xi$ for magnetic
		field $0.1\Hct$. The numbers in parentheses represent the volume
		fractions occupied by the particles $f$. The critical currents are
		determined by the intersection of the CVD and 2\% of the flux flow
		voltage (dashed line). The optimal concentration of particles is at
		$\Np = 4\,000$ corresponding to $f = 0.23$.
		(b)~The dependences of the critical current $\Jc$ on the particle
		volume fraction~$f$ for $a = 4\xi$ and three magnetic fields. The
		optimal~$f$ increases with the magnetic field.
		(c)~The dependences of~$\Jc$ on~$f$ for different particle
		diameters at $B = 0.1\Hct$. The optimal particle diameter is in
		between 3$\xi$ and 4$\xi$.
		(d)~Order parameter isosurfaces for a pinned vortex configurations
		with $a = 4\xi$, $\Np = 500$, and $B = 0.016\Hct$.
		(e)~The field-induced vortex lines extracted from the same order parameter.
		The particles are shown as transparent spheres. Vortex lines outside
		particles are red, and inside particles are blue.
	}
\end{figure*}

These limitations are overcome in the time-dependent Ginzburg-Landau (TDGL) model~\cite{Schmid:1966} which describes dynamics of the superconducting order parameter. In this description, the vortex lines appear spontaneously as singularities in the order parameter. Even though the TDGL model does not provide a fully realistic description of the dynamic properties of superconductors, it does describe accurately vortex-line flexibility, interactions between vortex lines, and interactions of vortices with pinning sites. It also allows for cuttings and reconnections of vortex lines. Therefore, this model is perfectly suited for the problem of the critical-current optimization, for which a fully accurate description of the dynamics is not essential. The TDGL model has been proven to be very useful for exploring many properties of the vortex state~\cite{Doria:1990,Machida:1993,Crabtree:1996,*BraunPhysRevLett96,*Crabtree:2000,Aranson:1996,SchweigertPhysRevLett98,*BaelusPhysRev02,WinieckiA:2002,BerdiyorovPhysRevB06,*Berdiyorov:2014,Vodolazov:2013}. However, only recently a meaningful exploration of the parameter space for sufficiently large three-dimensional (3D) superconductor has became possible~\cite{TDGLCode14} allowing us to address the problem of critical-current optimization.

We use the TDGL model to explore vortex pinning by randomly-placed metallic spherical inclusions. Our objective is to find the optimal parameters for the pinning landscape to maximize the critical current. Clearly, when the particles occupy a small fraction of the total volume, the critical current grows as particle density increases. At some density, however, further increase of particle number will not improve current-carrying capacity~\cite{GurevichAnnRevCMP14} due to at least two factors: (i)~The increasing mobility of the vortex lines due to jumping between the particles and (ii)~the reduction the effective cross section for the supercurrent caused by inclusions. Therefore, it is important to find the size and density of particles that maximizes the critical current. This problem can not be accessed by simple approaches. We find optimal parameters for different magnetic fields by systematically exploring the dependence of the critical current on size and concentration of particles.

The dynamics of the order parameter $\psi(\mathbf{r},t)$ is described by the TDGL equation in the reduced form
\begin{multline}
	(\partial_t + \imath\mu)\psi
	= \epsilon(\mathbf{r})\psi
	- |\psi|^2\psi \\
	+ \! \sum_{j=x,y,z} \!\! \eta_j^2(\nabla_j - \imath A_j)^2\psi
	+ \zeta(\mathbf{r},t).
	\label{eq:TDGL}
\end{multline}
Here $\mu$ and $\mathbf{A}$ are the scalar and vector potentials. We used the in-plane coherence length $\xi$ at the chosen temperature as the unit of length, meaning that $\eta_x = \eta_y = 1$ and $\eta_z = 1/\gamma$, were $\gamma$ is the anisotropy factor. We took $\gamma = 5$ corresponding to YBCO. The function $\epsilon(\mathbf{r})$ models pinning centers, $\epsilon(\mathbf{r}) = 1$ in the bulk~\footnote{In Ref.~\cite{TDGLCode14} a somewhat different normalization has been considered where $\epsilon(\mathbf{r})$ in the bulk was taken as $\epsilon = \Tc/T - 1$ and unit of length was the zero-temperature coherence length $\xi_{0}$. It is straightforward to demonstrate that selection of $\epsilon = 1$ is equivalent to transition to new units in which lengths are normalized to the temperature-dependent coherence length $\xi(T) = \xi_{0} / \sqrt{\Tc/T - 1}$.} and $\epsilon(\mathbf{r}) = -1$ inside metallic inclusions. We used the approximation of large London penetration depth $\lambda$ in which the vector potential $\mathbf{A}$ is fixed by the external magnetic field, $A_y = B x$, and the magnetic field is measured in units of the $c$-axis upper critical field at given temperature, $\Hct =\Phi_0/(2\pi\xi^2)$. The Langevin term $\zeta (\mathbf{r},t)$ describing thermal noise has the correlation function $\langle\zeta^*(\mathbf{r},t) \zeta(\mathbf{r}',t') \rangle = T \, \delta(\mathbf{r}\! -\! \mathbf{r}') \, \delta(t\! -\! t') $, where $T$ is the reduced temperature in units of $\Hc^2\xi^3/(8\pi)$ and $\Hc$ is the thermodynamic field. The total electric current in units of $J_0 = c\Phi_0/(8\pi^2\xi \lambda^2)$ (CGS) is given by
\begin{equation}
	J_j = \eta_j^2\left(\mathrm{Im} [ \psi^*(\nabla_j -\imath A_j)\psi ] - \nabla_j \mu\right).
	\label{eq:Jgen}
\end{equation}
where the first term is the supercurrent and the second term gives the normal current. In these units the depairing current is $\Jdp = 2/(3\sqrt{3}) J_0 \approx 0.385 J_0$. We performed simulations with fixed current applied in the $x$ direction and compute the average electric field $E = -\nabla_x \mu$ in the dynamic steady state. For the simulations, we developed a stable and efficient solver implemented for graphics processing units (GPUs) allowing for simulations of quite large 3D systems with sizes up to 500 coherence lengths in all directions~\cite{TDGLCode14}.

We systematically computed the current-voltage dependences (CVDs) for different particle sizes and densities. The simulated system size is $100\xi \! \times \! 100\xi \! \times \! 50\xi$ with $256 \! \times \! 256 \! \times \! 128$ mesh points. Figure~\subref{fig:IV} shows representative series of CVDs for different numbers of particles $\Np$ with diameter of $a = 4\xi$ at magnetic field $B = 0.1\Hct$ corresponding to 159 vortex lines in the system. These CVDs are obtained by stepwise decrease of the applied current with a simulation time of about $10^5$ TDGL units between current steps. Typically, we did not observe significant history effects: CVDs computed with different starting currents and current steps are found to be very similar to each other. Only at small magnetic fields $\sim 0.01 \Hct$ CVDs become more noisy and slightly history-dependent.

The pinning effectiveness of the particles is primarily determined by the volume fraction $f$ occupied by them. For spatially separated particles the ``nominal'' volume fraction is $\fn = \pi \Np a^3/(6L_xL_yL_z)$. As in our case randomly-placed particles may overlap, the real volume fraction is somewhat smaller, $f \approx \fn - \fn^2/2$, where the correction term accounts for possible overlaps between pairs of neighboring spheres. For each number of particles, the value of $f$ is specified in parenthesis in Fig.~\subref{fig:IV}. For low $f$ the CVDs systematically shift to the right as particle density increases indicating an increase of the critical current. Above a certain density adding more particles starts to degrade the critical current. The optimal density corresponds to the volume fraction $f = 0.23$.

\begin{figure}[b]
	\begin{center}
		\subfloat{\includegraphics[width=3.1in]{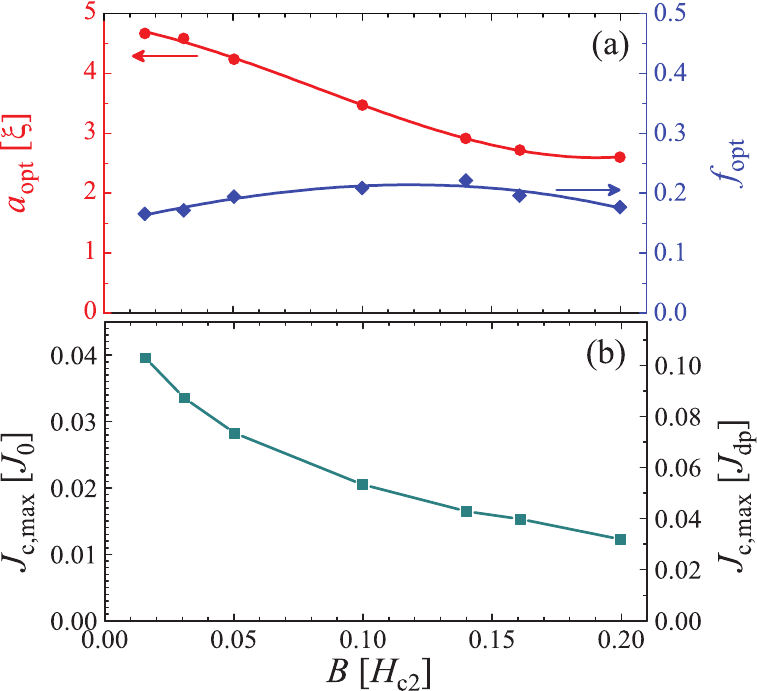} \label{fig:a_f_opt}}
		\subfloat{\label{fig:Jc_opt}}
	\end{center} \vspace{-0.2in}
	\caption{
		(a)~The magnetic field dependence of the optimal diameter
		(left axis) and volume fraction (right axis).
		(b)~The maximum critical current for optimal parameters for
		different magnetic fields in reduced units. The right axis shows
		this current normalized to the depairing current.
	}
\end{figure}

The computed CVDs are used to evaluate the critical currents, $\Jc$, which we define using the criterion $E(\Jc) = 0.02 \ff \Jc$ [dashed line in Fig.~\subref{fig:IV}]. The free flux-flow resistivity, $\ff$, in our reduced units is $\ff = 1.689 B$. Figure~\subref{fig:Jc_f_B} shows the dependences of the critical current on the nonsuperconducting volume fraction $f$ for $a = 4\xi$ and three magnetic fields. The optimal volume fraction slowly increases with the magnetic field, from $\sim 0.15$ at $B = 0.016\Hct$ to $\sim 0.23$ at $B = 0.1\Hct$. At the lowest field, $B = 0.016\Hct$, the maximum current is $\Jc \approx 0.0383 J_0$ corresponding to 10\% of the depairing current. The movie in~\footnote{\href{http://youtu.be/pPeMoZzKujY}{Animation} in supplemental materials illustrates evolution of vortex dynamics for $B = 0.016\Hct$ with increasing particle density.} demonstrates the evolution of vortex dynamics with increasing particle density for this magnetic field.

Figure~\subref{fig:Jc_f_a} presents the dependences of the critical currents $\Jc$ on the nonsuperconducting volume fraction $f$ for different particle diameters at $B = 0.1\Hct$. The highest critical current is realized at $a = 3\xi$ indicating the existence of an optimal particle size. By applying parabolic fits to the numerical data for the geometry dependent $\Jc(a,f)$, we found the optimal particle size and volume fraction at several fields. Figure~\subref{fig:a_f_opt} shows the magnetic field dependences of the optimal particle size $a_\mathrm{opt}$ and volume fraction $f_\mathrm{opt}$ in the range $0.016\Hct < B < 0.2\Hct$. Within this range the optimal size monotonically decreases with the increasing magnetic field from $\sim 4.5\xi$ to $\sim 2.5\xi$. This indicates that the typical scale of disorder has to be comparable with the intervortex spacing which decreases as $1/\sqrt{B}$. The optimal volume fraction has a weak, non-monotonic dependence on the field strength, but stays within the range of 17--22\%. The magnetic-field dependence of the maximum critical current achieved for the optimal parameters is shown in Fig.~\subref{fig:Jc_opt}.

To characterize the structure of the pinned vortex states, we extracted the field-induced vortex lines from the order parameter~\cite{PhillipsPRE2015} and performed a detailed analysis of these configurations. Figure~\subref{fig:isosurfaces} shows a representative trapped vortex configuration for $B = 0.016\Hct$ imaged by the order-parameter isosurfaces $|\psi(\mathbf{r})| = 0.1$. Both particles and vortex lines can be seen as regions of suppressed order parameter. We found that the vortex arrangements typically are quite disordered, which is partly caused by the interaction of the vortices with randomly arranged particles and partly by incomplete equilibration. Vortex lines traced from this order-parameter distribution are shown in Fig.~\subref{fig:vortices}. They are split into line segments located inside the particles and in superconducting material, as illustrated by blue and red lines, respectively.

\begin{figure}[tb]
	\begin{center}
		\subfloat{\includegraphics[width=3.00in]{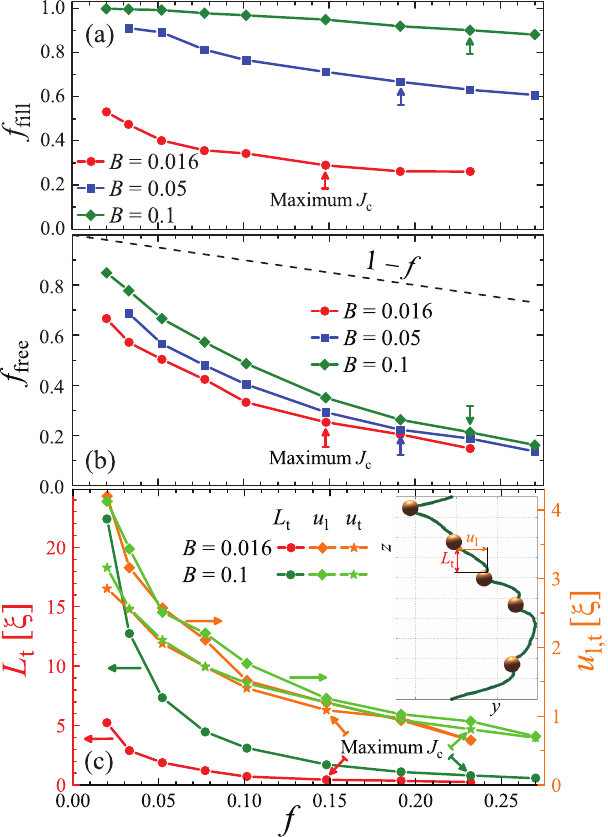} \label{fig:f_fill}}
		\subfloat{\label{fig:f_free}}
		\subfloat{\label{fig:Lu}}
	\end{center} \vspace{-0.2in}
	\caption{
		Evolution of parameters characterizing pinned vortex-line
		configurations with increasing nonsuperconducting volume
		fraction $f$ for $a = 4\xi$ and three magnetic fields.
		The plot (a) shows the fraction of particles occupied by vortex
		lines, $\ffill$.
		The plot (b) presents the length fraction of vortex segments
		outside the particles, $\ffree$. Dashed line shows
		the volume fraction occupied by superconducting material,
		$1 \! - \! f$. The plot (c) shows the average geometrical parameters
		characterizing pinned line segments as illustrated in the inset,
		the segment length $\Lt$ (left axis) and typical
		pin-to-pin line displacements in the direction of motion,
		$\ul$, and in the transverse direction, $\ut$
		(right axis). The parameters $\Lt$ and $\ul$
		are defined in the inset. In all plots arrows mark locations of the
		maximum critical current.
	}
\end{figure}

We extracted several parameters characterizing trap\-ped configurations: (i)~The fraction of particles occupied by vortices, $\ffill$, (ii)~the fraction of particles double-occupied by vortices, $f_2$, (iii)~the fraction of the total line length located outside particles, $\ffree = \ell_\mathrm{outside} / \ell_\mathrm{total}$, (iv)~the average length of line segments trapped between neighboring particles, $\Lt$, see inset in Fig.~\subref{fig:Lu}, and (v)~the average particle-to-particle displacement in the direction of motion, $\ul = u_y$, and in transverse direction $\ut = u_x$. These parameters are not independent. Indeed, the number of particles holding a given vortex line can be estimated as $\ffree L_z / \Lt$ meaning that the total number of occupied pins is $\Nv \ffree L_z / \Lt$, where $\Nv$ is the total number of vortex lines. As fraction of pins $f_2$ is holding two vortex lines, we can estimate the occupied fraction as $\ffill \approx \Nv \ffree L_z / [\Np(1+f_2) \Lt]$. We checked that the extracted parameters satisfy this consistency condition.

Particle fraction occupied by the vortex lines, $\ffill$, naturally characterizes the efficiency of pin utilization. Figure~\subref{fig:f_fill} shows the dependence of this parameter on the nonsuperconducting volume fraction $f$ for $a = 4\xi$ and three magnetic fields. As expected, this parameter increases with the magnetic field and decreases with the number of particles. For $B = 0.1\Hct$ almost all particles are occupied. We also found that for this field typically 3--5\% of particles hold two vortex lines without any systematic dependence on the particle density. For smaller fields the number of double-occupied particles is negligible. An noteworthy feature is that for small fields a significant fraction of particles remains unoccupied even for very low particle densities.

The efficiency of the vortex trapping by the particles can be characterized by the free-segment length fraction of the vortex lines, $\ffree$, plotted in Fig.~\subref{fig:f_free}. A natural upper limit for $\ffree$ is the volume fraction occupied by the superconducting material, $1 - f$, shown by the dashed line. This limit would be realized if there were no correlations between particles and vortices. We can see that for pinned configurations $\ffree$ is significantly below this limit since particles trap vortices. Notably, $\ffree$ drops below 50\% when particles occupy only 5\% of the volume for $B = 0.016\Hct$. The free-segment fraction monotonically increases with the magnetic field, because the vortex lines compete for the particles. We observe that the maximum critical current is realized for $\ffree = 21$--25\% which only weakly depends on the magnetic field.

For strong pinning sites the vortex lines split into finite-size segments hanging in between neighboring sites~\cite{OvchinnikovI:1991,BlatterGK:2004,Koshelev:2011}. Figure~\subref{fig:Lu} shows the behavior of the average length parameters in units of $\xi$ characterizing these free-line segments, the segment length $\Lt$ and pin-to-pin line displacements in the direction of motion, $\ul$, and in the transverse direction, $\ut$. The length $\Lt$ rapidly increases with decreasing particle density and with increasing magnetic field. While the displacements $u_{\mathrm{l,t}}$ also increase with decreasing $f$, they only weakly depend on the magnetic field. As expected, for small particle densities the vortices stretch between pinning sites preferentially in the direction of motion~\cite{OvchinnikovI:1991,BlatterGK:2004,Koshelev:2011} meaning that $u_l$ is larger than $u_t$. These parameters, however, become almost identical for $f \gtrsim 0.15$. At the particle density corresponding to the maximum of the critical current all three parameters, $\Lt$, $\ul$, and $\ut$, are close to $\xi$ for both magnetic fields.

In summary, we investigated vortex pinning by randomly distributed metallic inclusions and found optimal parameters for highest critical current. We also analyzed statistical properties of pinned vortex arrays and revealed several nontrivial structural properties of the optimally-pinned states. Our general observation is that there is no universal optimal pinning configuration for all magnetic fields. Thus, for best performance in a given application, pinning landscapes should be designed taking into account relevant magnetic fields.

The authors acknowledge fruitful discussions with W.\,K.\,Kwok, U.\,Welp, M.\,Leroux, V.\,B.\,Geshkenbein, and R.\,Willa. This work was supported by the Scientific Discovery through Advanced Computing (SciDAC) program funded by U.S. DOE, Office of Science, Advanced Scientific Computing Research and Basic Energy Science. A.E.K. was supported by the Center for Emergent Superconductivity, an Energy Frontier Research Center funded by the U.S. DOE, Office of Science, Office of Basic Energy Sciences. C.L.P. was funded by the Office of the Director through the Named Postdoctoral Fellowship Program (Aneesur Rahman Postdoctoral Fellowship), Argonne National Laboratory.

\bibliography{spherical}

\end{document}